\documentclass{article}
\PassOptionsToPackage{numbers}{natbib}
\usepackage[preprint]{neurips_2019}
\usepackage[utf8]{inputenc} 
\usepackage[T1]{fontenc}    
\usepackage{hyperref}       
\usepackage{url}            
\usepackage{booktabs}       
\usepackage{amsfonts}       
\usepackage{nicefrac}       
\usepackage{microtype}      
\usepackage{graphicx}
\usepackage{subcaption}
\usepackage[justification=centering]{caption}
\usepackage{multirow}
\usepackage{graphicx}

\title{Clusters in Explanation Space: Inferring disease subtypes from model explanations}

\author{
Marc-Andre Schulz\thanks{Corresponding Author} \\
Department of Psychiatry, Psychotherapy\\ and Psychosomatics \\ RWTH Aachen University \\ Aachen, Germany \\
\texttt{marc.schulz@rwth-aachen.de} \\
\And
Matt Chapman-Rounds \\
School of Informatics \\ University of Edinburgh \\ Edinburgh, United Kingdom \\
\texttt{m.rounds@ed.ac.uk} \\
\And
Manisha Verma \\
QuantumBlack \\ London, United Kingdom \\
\texttt{manisha.verma@quantumblack.com}
\And
Danilo Bzdok \\
Department of Biomedical Engineering \\
Faculty of Medicine \\
McGill University \\
Montreal, Canada \\
\texttt{danilo.bzdok@mcgill.ca}
\And
Konstantinos Georgatzis \\
QuantumBlack \\ London, United Kingdom \\
\texttt{konstantinos.georgatzis@quantumblack.com}
}
 
\begin{document}

\maketitle

\begin{abstract}

Identification of disease subtypes and corresponding biomarkers can substantially improve clinical diagnosis and treatment selection. Discovering these subtypes in noisy, high dimensional biomedical data is often impossible for humans and challenging for machines.

We introduce a new approach to facilitate the discovery of disease subtypes: Instead of analyzing the original data, we train a diagnostic classifier (healthy vs. diseased) and extract instance-wise explanations for the classifier's decisions.
The distribution of instances in the \textit{explanation space} of our diagnostic classifier amplifies the \textit{different reasons for belonging to the same class} - resulting in a representation that is uniquely useful for discovering latent subtypes.

We compare our ability to recover subtypes via cluster analysis on model explanations to classical cluster analysis on the original data. In multiple datasets with known ground-truth subclasses, most compellingly on UK Biobank brain imaging data and transcriptome data from the Cancer Genome Atlas, we show that cluster analysis on model explanations substantially outperforms the classical approach.

While we believe clustering in explanation space to be particularly valuable for inferring disease subtypes, the method is more general and applicable to any kind of sub-type identification.

\end{abstract}

\pagebreak

\section{Introduction}
Many diseases manifest differently in different humans. This heterogeneity is especially pronounced in fields like psychiatry or oncology where visible symptoms are far removed from the  underlying pathomechanism. What appears to be a coherent collection of symptoms is often the expression of a variety of distinct disease subtypes, each with different disease progression and different treatment responses. Dealing with these subtypes is the purview of precision medicine: identified subtypes and their corresponding biomarkers are used to refine diagnoses, predict treatment responses and disease progression, and to inform further scientific research \citep{bzdok2017machine}.

The search for biologically grounded subtypes is frequently carried out by means of cluster analysis to identify groups of subjects with similar disease phenotypes. Cluster analysis is applied in either the original set of variables or feature space (e.g. voxels of a brain MRI) or in some hand-crafted aggregate measures inspired by existing external knowledge (e.g. gray matter density of predefined brain regions)  \citep{carey2006race, erro2013heterogeneity, drysdale2017resting}. Identifying clusters is particularly challenging when faced with high dimensional feature spaces, such as MRI or genome data, where the high dimensionality and a generally low signal to noise ratio impede the straightforward application of modern clustering algorithms. Furthermore, similarity in the original feature space (e.g. brain MRI) is not necessarily informative about the investigated disease. That is, the most obvious sources of variation useful for groupings might reflect sex or age instead of similar disease phenotypes.  Thus, biomedical data in particular often needs to be transformed from the original feature space into a more informative embedding space.  In this paper, we propose a novel space that we believe to be particularly useful for identifying latent subtypes: the space of explanations corresponding to a diagnostic classifier.

Recent interest in explaining the output of complex machine learning models has been characterized by a wide range of approaches \citep{lipton2016mythos, MontavonSM17}, most of them focused on providing an instance-wise explanation of a model’s output as either a subset of input features \citep{anchors:aaai18, chen2018learning}, or a weighting of input features \citep{ribeiro2016should,lundberg2017unified}. The latter, where each input feature is weighted according to its contribution to the underlying model’s output for an instance, can be thought of as specifying a transformation from \textit{feature space} to an \textit{explanation space}. This explanation space is conditioned on the underlying model’s output. Simply put, every sample that can be classified by a model has a corresponding explanation. This explanation consists of the contributions of each individual input feature to the classification result. The explanation thus has the same dimensionality as the original feature space. This space of feature contributions (the explanation space) can be thought of as a new feature space in its own right and can itself serve as a basis for classification. In contrast to the original feature space, the new explanation space directly relates to the classification goal. In the case of a diagnostic classifier (healthy vs. diseased), the explanation space relates to the investigated disease.

We argue that the explanation space of a diagnostic classifier is an appropriate embedding space for subsequent cluster analyses aimed at the discovery of latent disease subtypes.
Firstly, the explanations should collapse features that are irrelevant to the classification of a particular disease, thereby mitigating the Curse of Dimensionality. Secondly, we expect different disease subtypes to have structurally different explanations for belonging to a disease class. The explanation space of a diagnostic classifier is expected to amplify the \textit{different reasons for belonging to the same class} - resulting in a representation that should be uniquely useful for recovering latent subtypes. The intuition here is that instance-wise explanations refer to a \textit{local} part of the classifier's decision boundary. This means that if the decision boundary is differently oriented for different parts of the space (due, for example, to multiple distinct underlying subclasses), the explanations will also differ meaningfully.

For a proof of principle, we take the approach of converting multi-class classification problems into binary ones for the purposes of training the underlying model. This means that each ‘class’ has several distinct subclasses, of which the classifier is unaware - while we retain ground-truth knowledge of the respective subclasses. In four revisited datasets, we demonstrate that the clusters in explanation space recover these known subclasses.

\section{Related Work}

Examples of clustering work in the original feature space include \citet{carey2006race}, who use a hierarchical clustering approach to population-based distributions and clinical associations for breast-cancer subtypes, and \citet{erro2013heterogeneity}, who use a k-means based clustering approach to test the hypothesis that the variability in the clinical phenotype of Parkinson’s disease was caused by the existence of multiple distinct subtypes of the disease. \citet{drysdale2017resting} show that patients with depression can be subdivided into four neurophysiological subtypes, by hierarchical clustering on a learned embedding space.

Using instance-wise explanations for clustering has previously been discussed in \citet{lundberg2018consistent}, under the name “supervised clustering”. These authors show that clustering on Shapley values explains more model variance than other tree-specific feature importance estimates \citep{lundberg2018consistent, saabas2014}. To the best of our knowledge, no previous work has shown if, and to what extent, clusters in explanations space can be of practical relevance, nor has any work drawn the link to disease-subtyping in precision medicine. The present work bridges the gap between recent advances in \textit{explainable AI} and real-world medical applications, by linking the former to a crucial biomedical problem and providing not only a proof-of-concept, but a full, clinically relevant, example - the discovery of cancer-subtypes.

\section{Data}

We chose four datasets of increasing complexity to evaluate the efficacy of subclass recovery in explanation space. Firstly, simple synthetic data for a proof-of-concept. Secondly, Fashion-MNIST as a machine learning benchmark. Thirdly, age prediction from brain imaging data as a simple biomedical example. Lastly, cancer subtype detection as a challenging real-world biomedical problem. Brain imaging data was provided by the UK Biobank, cancer transcriptome data from the Cancer Genome Atlas. The two are among the world's largest biomedical data collections and represent the two most likely fields of applications: precision psychiatry and oncology, with \textit{big data} in both \textit{p} (dimensionality) and \textit{n} (sample size).

The original Madelon dataset from the 2003 NIPS feature selection challenge \citep{guyon2005result} is a synthetic binary classification problem, with data points placed in clusters on the vertices of a hypercube in a subspace of the feature dimension. The scikit-learn implementation \citep{scikit-learn} of the generative algorithm was used to create a variation with 16 classes, 50 features, and data points distributed in one cluster per class on the vertices of a four-dimensional hypercube. 

Zalando provides Fashion-MNIST \citep{xiao2017fashion} as a more challenging drop-in replacement for MNIST, with the same dimensionality (782) and sample size (70,000). Instead of the 10 classes of digits (0-9) of the original MNIST dataset, Fashion-MNIST consists of grayscale images of ten classes of clothing (e.g. T-shirts, pants, dresses). We chose Fashion-MNIST, because the original MNIST has been argued to be too easy a problem for modern methods. Indeed, most classes in the dataset can already be separated in the first two principal components of the feature space, making it ill-suited for the present analysis.

The UK Biobank (UKBB) \citep{miller2016multimodal} is one of the world’s largest biomedical data collections. It provides, amongst a multitude of other phenotypes, structural (T1) brain MRI data for 10000 participants. The structural brain MRI images were preprocessed into 164 biologically motivated imaging-derived phenotypes (IDPs), prepresentin aggregated grey and white matter densities per brain region. For a biomedical proof-of-concept, we chose age as simple target variable - cut into 4 quartile "classes".

The Cancer Genome Atlas (TCGA) project \citep{weinstein2013cancer} provides transcriptome\footnote{The transcriptome consists of all “transcripts”, i.e.\ copies, of DNA into RNA, that are necessary to implement DNA instructions.} data for various forms and subtypes of cancer. The data consists of 60,498 gene expression (fpkm) values for 8500 participants (after removing participants with missing values). We chose the cancer tissue's immune model based subtype (6 classes, i.e.\ Wound Healing, IFN-gamma Dominant, Inflammatory, Lymphocyte Depleted, Immunologically Quiet, TGF-beta Dominant) as a complex, clinically relevant target variable.

Our synthetic and Fashion-MNIST datasets have balanced classes by design. The UK Biobank brain imaging data is cut into 4 age-quartiles resulting in balanced classes. The TCGA dataset is imbalanced. For classification it was sampled inversely proportional to class frequencies. 

\section{Methods}

For each dataset, classes were split into two supersets, one containing the first three\footnote{The first and fourth class for UKBB.} classes, the other containing the remainder \footnote{The decision to split into three-vs-rest was due to ease of visualization, and simplicity of choice.}. This yielded a new binary classification problem. This setup allows for evaluating approaches to subtype discovery. The binary classification data is intended to represent observable characteristics (e.g.\ the healthy vs. diseased distinction in medicine) while the original classes represent the hidden subtypes. 

A Random Forest classifier \citep{breiman2001random} was trained on the binary classification problem and SHapley Additive Explanations (SHAP) \citep{lundberg2017unified} were used to generate instance-wise explanations for the predictions of the classifier. 
Approaches to model agnostic explanations tend to be highly computationally expensive, as they typically use extensive sampling procedures to estimate local approximations of decision boundaries or to find proximate counterfactual instances. The specific combination of classification model and explanation approach motivated by the fact that there exists a particularly efficient solution to the computation of Shapley values in the case of Random Forests \citep{lundberg2018consistent}.
The concept of Shapley values [21] has its origins in cooperative game theory. Shapley values indicate how to distribute payoffs of a cooperative game proportional to each player’s individual contribution. SHAP interprets prediction as a cooperative game and assigns each input variable or feature its marginal contribution to the predicted outcome. Investigating the applicability of other explanation methods and classifiers to specific domains would be a natural direction for future work.

Each explanation is represented as a vector of the same dimensionality as the original feature space, indicating how strongly and in what direction each feature contributed to the prediction result of a given observation. The space spanned by the model explanations was then compared to the original feature space, with respect to the distinguishability of ground-truth subtypes (the original classes of the dataset).

\section{Results}

\begin{figure}[ht]
    \centering
    \includegraphics[width=\linewidth]{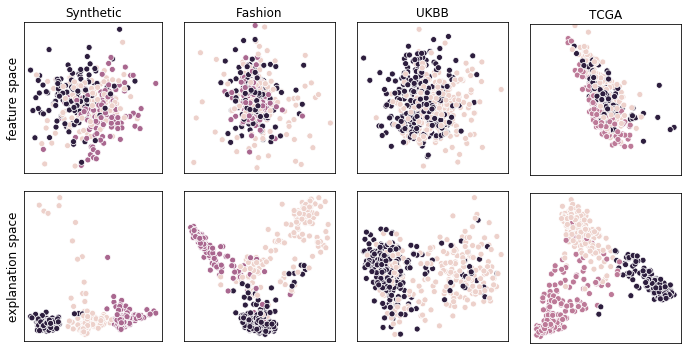}
  \caption{
Projection into first two principal components after PCA of feature space (top row) and explanation space (bottom row) of Synthetic, Fashion-MNIST, UK Biobank Brain-Age, and Cancer Genome Atlas datasets. Shown are the (sub-)classes of the three-class superset used for binary classification. As can be seen, clusters are well defined in explanation space; for biomedical datasets, these sub classes correspond to real subtypes. For example in TCGA (rightmost plot), the visible clusters correspond to Wound Healing (black), IFN-gamma Dominant (yellow), and Inflammatory (pink) cancer tissue immune model types.}
    \label{pca}
\end{figure}

To compare the correspondence of clustering in the respective spaces to the underlying distribution of class labels we used three qualitatively distinct approaches. Firstly, we visually inspected the projection into the first two principal components. Distinct clusters in the data would constitute major directions of variance and should be easily visible in the PCA projection, allowing for a first sanity check (Fig. \ref{pca}).
Secondly, we evaluated standard clustering quality indices to quantify structural differences between representations. We chose the Davies-Bouldin index, defined as the average similarity between each cluster $C_i$ and its most similar other cluster $C_j$ \citep{davies1979cluster}, the Silhouette Coefficient, which balances the mean distance between a sample and all other points in its cluster with the mean distance between that sample and all the points in the next nearest cluster \citep{rousseeuw1987silhouettes}, and the Calinski-Harabaz Index, which is given as the ratio of the between-clusters dispersion mean and the within-cluster dispersion \citep{calinski1974dendrite}.
Lastly, to ensure that transformation from feature space into explanation space really does lead to improved subclass recovery, we applied Agglomerative Clustering and report the  Adjusted Mutual Information \citep{vinh2010information} between reconstructed subclasses with ground-truth labels.

To compare improvements derived from transformation into explanation space with those which can be gained by dimensionality reduction alone, we apply\footnote{All run with scikit-learn default parameters.} PCA \citep{friedman2001elements}, Isomap \citep{tenenbaum2000global}, and T-SNE \citep{maaten2008visualizing} to both feature space and explanation space, reducing their dimensionality down to two latent factors. These dimensionality reduction methods were selected to represent a diversity of approaches. PCA is strictly linear, while Isomap and T-SNE allow for non-linear interactions. In contrast to Isomap and T-SME, PCA imposes orthogonality on latent factors. While PCA focusses on closeness in input-space, Isomap and T-SNE optimize closeness in embedding space. 
We report clustering quality indices and Mutual Information for both feature space and explanation space in the original dimensionality as well as the reduced PCA, Isomap, and T-SNE spaces.

Quantitative results are shown in Table \ref{score_table}. Clustering quality indices and post-clustering Mutual Information consistently improved (average  AMI gain of 0.45) when moving from feature space to explanation space in all four datasets, both before and after dimensionality reduction. Dimensionality reduction improved subclass recovery in both feature space and explanation space, with PCA performing worst, T-SNE performing best, and Isomap somewhat inconsistently in-between. Notably, improvement derived from dimensionality reduction was larger when reducing explanation space than when reducing feature space, with an average AMI gain from T-SNE of 0.06 in feature space and 0.13 in explanation space.

\section{Conclusion}

We propose \textit{explanation space} as a powerful embedding to facilitate detection of latent disease subtypes, particularly by cluster analysis. In four revisited datasets, we have shown both that the distribution of data points in explanation space is sharply clustered, and that these clusters more accurately correspond to ground-truth subclasses than clusters derived from the original feature space.

We also demonstrated the relevance of our approach to the real world problem of disease subtype discovery. In both the Cancer Genome Atlas and the UK Biobank brain imaging data, task-conditioned explanations proved to be highly informative - to the extent that subtypes can be easily visually identified (Fig. \ref{pca}).

Existing approaches to subtype discovery \citep{carey2006race,erro2013heterogeneity,drysdale2017resting} can easily be adapted to work on task conditioned explanations. Disease status labels are naturally available and can be used to train a diagnostic classifier. The classifier's instance-wise explanations have the same dimensionality as the original data and can serve as a drop-in replacement.

Transforming from feature space to explanation space should not be seen in competition to dimensionality reduction, but rather as a complimentary processing step. Both help to recover latent subtypes  and the benefits from dimensionality reduction appear to be substantially amplified by first transforming into an appropriate explanation space.

We hope that this work will serve as a starting point for further exploration of explanation-based
methods for inferring disease subtypes. More broadly, the fact that off-the-shelf explanatory tools can be used to generate task-specific embeddings is undoubtedly a promising avenue for a variety of applications.

\pagebreak

\begin{table}[hbt!]
\centering
\resizebox{\textwidth}{!}{%
\begin{tabular}{lll|rrrr}
Data                      & Dim. Red.               & Space & \multicolumn{1}{l}{Davies-Bouldin} & \multicolumn{1}{l}{Calinski-Harabaz} & \multicolumn{1}{l}{Silhouette} & \multicolumn{1}{l}{Adjusted MI} \\ \hline
\multirow{8}{*}{Sythetic} & \multirow{2}{*}{none}   & (f)   & 7.00                               & 6.52                                 & 0.02                           & 0.08                            \\
                          &                         & (e)   & 1.11                               & 192.62                               & 0.46                           & 0.59                            \\ \cline{2-7} 
                          & \multirow{2}{*}{PCA}    & (f)   & 3.30                               & 38.08                                & 0.01                           & 0.14                            \\
                          &                         & (e)   & 0.78                               & 258.24                               & 0.66                           & 0.59                            \\ \cline{2-7} 
                          & \multirow{2}{*}{Isomap} & (f)   & 12.90                              & 3.93                                 & -0.02                          & 0.02                            \\
                          &                         & (e)   & 0.72                               & 187.73                               & 0.71                           & 0.52                            \\ \cline{2-7} 
                          & \multirow{2}{*}{T-SNE}  & (f)   & 29.45                              & 0.61                                 & -0.02                          & 0.00                            \\
                          &                         & (e)   & 0.29                               & 2937.57                              & 0.79                           & 0.97                            \\ \hline
\multirow{8}{*}{Fashion}  & \multirow{2}{*}{none}   & (f)   & 11.10                              & 2.86                                 & -0.03                          & 0.00                            \\
                          &                         & (e)   & 1.32                               & 205.45                               & 0.30                           & 0.61                            \\ \cline{2-7} 
                          & \multirow{2}{*}{PCA}    & (f)   & 5.72                               & 2.18                                 & -0.12                          & 0.00                            \\
                          &                         & (e)   & 0.74                               & 581.87                               & 0.51                           & 0.56                            \\ \cline{2-7} 
                          & \multirow{2}{*}{Isomap} & (f)   & 4.87                               & 75.60                                & -0.07                          & 0.00                            \\
                          &                         & (e)   & 0.92                               & 398.96                               & 0.41                           & 0.56                            \\ \cline{2-7} 
                          & \multirow{2}{*}{T-SNE}  & (f)   & 2.30                               & 65.03                                & 0.12                           & 0.13                            \\
                          &                         & (e)   & 0.69                               & 768.13                               & 0.52                           & 0.71                            \\ \hline
\multirow{8}{*}{UKBB}     & \multirow{2}{*}{none}   & (f)   & 4.24                               & 26.38                                & 0.04                           & 0.04                            \\
                          &                         & (e)   & 2.33                               & 84.91                                & 0.12                           & 0.15                            \\ \cline{2-7} 
                          & \multirow{2}{*}{PCA}    & (f)   & 2.53                               & 57.65                                & 0.07                           & 0.05                            \\
                          &                         & (e)   & 1.54                               & 152.25                               & 0.21                           & 0.14                            \\ \cline{2-7} 
                          & \multirow{2}{*}{Isomap} & (f)   & 2.60                               & 60.75                                & 0.08                           & 0.04                            \\
                          &                         & (e)   & 1.74                               & 122.90                               & 0.21                           & 0.14                            \\ \cline{2-7} 
                          & \multirow{2}{*}{T-SNE}  & (f)   & 2.64                               & 59.99                                & 0.08                           & 0.05                            \\
                          &                         & (e)   & 1.44                               & 180.59                               & 0.24                           & 0.12                            \\ \hline
\multirow{8}{*}{TCGA}     & \multirow{2}{*}{none}   & (f)   & 8.58                               & 8.54                                 & 0.01                           & 0.10                            \\
                          &                         & (e)   & 2.64                               & 57.37                                & 0.13                           & 0.68                            \\ \cline{2-7} 
                          & \multirow{2}{*}{PCA}    & (f)   & 4.44                               & 17.81                                & -0.01                          & 0.05                            \\
                          &                         & (e)   & 0.59                               & 929.48                               & 0.59                           & 0.67                            \\ \cline{2-7} 
                          & \multirow{2}{*}{Isomap} & (f)   & 8.27                               & 70.70                                & 0.02                           & 0.16                            \\
                          &                         & (e)   & 0.51                               & 834.70                               & 0.62                           & 0.70                            \\ \cline{2-7} 
                          & \multirow{2}{*}{T-SNE}  & (f)   & 5.03                               & 106.36                               & 0.11                           & 0.28                            \\
                          &                         & (e)   & 0.51                               & 1561.08                              & 0.60                           & 0.73                           
\end{tabular}%
}
\caption{Cluster tightness and separation (Davies-Bouldin, Calinski-Harabaz, Silhouette), as well as subclass recovery performance (Adjusted Mutual Information after hierarchical clustering), in explanation space (e) and feature space (f), in original dimensionality and post PCA, Isomap, and T-SNE. Higher values are better (except for Davies-Bouldin, where lower is better).}
\label{score_table}
\end{table}

\pagebreak

\bibliography{c_explans}

\end{document}